\documentclass{skaox2006}
\begin{document}
   \title{Neutron Star Astronomy with the E-ELT\thanks{}}

   \author{R.P. Mignani\inst{1}}

   \institute{Mullard Space Science Laboratory, University College London, RH56NT, Dorking, Surrey, UK        }

   \abstract{So far, 24 Isolated neutron stars (INSs) of different types have been identified at optical wavelengths, from the classical radio pulsars to more peculiar objects, like the magnetars. Most identifications have been obtained in the last 20 years thanks to the deployment of modern technology telescopes, above all the {\em HST},  but also the {\em NTT}  and, later,  the 8m-class telescopes like the {\em VLT}.  The larger identification rate has increased the impact factor of optical observations in the multi-wavelength approach to INS astronomy, opening interesting possibilities for studies not yet possible at other wavelengths.  With the {\em HST} on the way to its retirement, 8m class telescopes will have the task of bridging neutron star optical astronomy into a new era, characterised by the advent of the generation of extremely large telescopes (ELTs), like the European ELT ({\em E-ELT}).  This will mark a major step forward in the field, enabling one to identify many more INSs, many of which from follow-ups of observations performed with future radio and X-ray megastruscture facilities like {\em SKA} and {\em IXO}. Moreover, the {\em E-ELT} will make it possible to carry out observations, like timing, spectroscopy, and polarimetry, which still represent a challenge for 8m-class telescopes and are, in many respects, crucial for studies on the structure and composition of the neutron star interior and of its magnetosphere.  In this contribution, I briefly summarise the current status of INS optical observations, describe the main science goals for the {\em E-ELT}, and their impact on neutron star physics. }
   
   \maketitle
%
%
\section{Introduction}

Optical observations of isolated neutron stars (INSs) have been performed soon after the discovery of the first radio pulsars. After the identification of the Crab pulsar in 1969 and, a few years later, of the Vela pulsar, both detected as optical pulsars, optical observations of INSs settled. At the beginning of the 1980s,  high-energy observations pinpointed interesting targets in newly discovered INSs which where  either intrinsically radio-faint, like PSR\, B0540$-$69 in the LMC, or radio-silent, like Geminga whose study greatly benefited from the help of optical astronomy.  In particular, the multi-wavelength chase for Geminga triggered follow-up optical observations of other  known X-ray and/or $\gamma$-ray detected pulsars, while optical observations were performed to ascertain the nature of some enigmatic high-energy sources suspected to be members of new classes of INSs, like the soft $\gamma$-ray repeaters (SGRs), the anomalous X-ray pulsars (AXPs), the X-ray Dim INSs (XDINSs), and the central compact objects (CCOs) in SNRs, or of peculiar transient radio pulsars  (see Kaspi 2010 for a recent review of all known INS types).  Thanks to the advent of modern technology facilities, like the ESO {\em NTT} (Mignani et al. 2000),  of the {\em HST} (Mignani 2010a), and of the 8m-class telescopes, like the {\em VLT} (Mignani 2009a),  the large observational effort resulted in many more INS optical identifications and paved the way to the first studies of the INS optical emission properties. So far, 24 INSs of different types  have been identified at optical wavelengths, albeit at different level of confidence (see Mignani 2009b, 2010b for a complete summary), including 12 rotation-powered pulsars, 7 magnetars, and 5 XDINSs.  
In this contribution,  I will describe the future role of the {\em E-ELT} in neutron star astronomy, building up on the {\em VLT} work (Section 2), and I will discuss some of the several scientific implications of optical observations of INSs on the understanding of the neutron star physics, formation, and evolution  (Section 3).


\section{The role of the E-ELT in neutron star astronomy}

Clearly, the major limitation in optical studies of INSs is represented by their intrinsic faintness (with fluxes down to $V\sim 27.5$) which requires a large collecting power. With a 42m mirror, the largest among all future ELTs, the {\em E-ELT} will provide a factor of $\sim 30$ improvement in collecting power with respect to the {\em VLT}, allowing to push the detection limit down to $V \approx 32$, i.e. almost two orders of magnitude in flux.  This will allow to detect many more INSs, so far too faint to be detected by the {\em VLT}  or barely detectable, up to larger distances and, thanks to the use of Adaptive Optics (AO), in the crowded and high-extinctIon regions of the galactic plane, e.g. where most magnetars are discovered. Thus, the {\em E-ELT} will be able to  reduce the current gap  between optical and high-energy observations,  where the number of detections in the X-ray and $\gamma$-ray bands amount to $\sim 80$ (Becker 2009) and $\sim 60$ (Abdo et al. 2010), respectively, with the radio detections ($\approx 1800$) clearly outnumbering those obtained in all the other energy bands.

Apart from  the currently known INSs,  many new targets will come from observations performed with the new generation of megastructure facilities, in particular from radio surveys performed with the {\em SKA} which is expected to discovered $\approx 30\,000$ new radio pulsars.  Thus, the privileged location of the {\em E-ELT} in the southern hemisphere (Cerro Armazonas, Chile), where the {\em SKA} will be built (Australia/South Africa), offers a unique optical/radio synergy chance. Indeed, radio pulsars still represent the largest fraction of the INS population and, thus, newly discovered objects will be natural  targets for {\em E-ELT} observations. Moreover, radio observations  give INS positions with unrivalled accuracy, distances through the dispersion measure (DM) or radio parallax, age and rotational energy loss rate $\dot{E}$ from the period $P$ and its first derivative $\dot{P}$, which are crucial to infer expectation values for the optical luminosity. The {\em E-ELT} will be able to  identify  $\approx 100$ among the known radio pulsars through direct imaging and to measure their flux with unprecedented accuracy for objects this faint.  In particular, the {\em E-ELT}, if equipped with high-time resolution instruments,   will allow to detect Crab-like optical pulsars in M31 (Shearer 2010).  The identification score will be large enough to  perform population studies of the optically-identified INSs and to investigate, on a much larger statistics,  the dependence of their optical emission properties on the pulsar parameters (e.g., age, period,  magnetic field).  Future targets will also come from X-ray observations performed with {\em IXO}. In particular, the combination of optical/IR and X-ray observations performed with the {\em E-ELT} and {\em IXO} will be crucial to investigate the nature of peculiar radio-silent INSs,  as done so far with the {\em NTT} and {\em ROSAT} in the 1990s and the {\em VLT} and {\em XMM-Newton} in the 2000s for the magnetars, the XDINSs, and the CCOs.  Finally, other targets will be provided {\it a posteriori} from mining the larger and larger INS database currently being built from $\gamma$-ray observations performed with {\em Fermi}, extremely efficient in spotting radio-silent  $\gamma$-ray pulsars  ($\sim 20$ and counting), like Geminga, whose study in the optical/IR might await for the {\em E-ELT}.  Deep imaging with the {\em E-ELT} will also allow to search for extended structures around INSs, associated either to pulsar-wind nebulae (PWNe) or to bow-shocks produced by the neutron star supersonic motion in the ISM, of which only 6 cases have been detected so far despite of several deep surveys with the {\em NTT} (Pellizzoni et al. in preparation). 

In principle, the large collecting area of the {\em E-ELT} will also make it possible to perform more detailed studies on the optically-identified INSs, including timing, spectroscopy, and polarimetry, which, in most cases,  are still a challenge for 8m-class telescopes. So far, these measurements have been obtained only for a handful  of objects out of the 24 INSs currently identified (see Table 1 of Mignani 2010b for an overview). In particular, optical  pulsations have been measured for 8 INSs only: 5 rotation-powered pulsars and 3 magnetars (see Mignani 2010b for an updated summary).  The {\em E-ELT} has the potential to increase this score substantially, making optical pulsars no longer a rarity (see, e.g. Shearer et al. 2008a,b). Medium-resolution optical or IR spectra have been obtained only for 8 INSs, albeit with very much different signal-to-noise levels, and an high-quality one has been obtained only for the Crab pusar. In most cases, this has left the knowledge of the INS optical spectrum very uncertain and based on sparse, and non homogeneous,  multi-band imaging photometry.  Finally, measurements of the phase-averaged polarisation have been obtained only for 2 INSs (with the Crab pulsar being the only one for which we have measurements of the phase-resolved polarisation across the whole period), with upper limits obtained for other 6. Only for the two brightest INSs, the Crab pulsar and PSR\, B0540$-$69,  measurements of the optical pulsations, spectrum, and polarisation exist.  It is clear that, in order to ensure a qualitative improvement in timing, spectroscopy, and polarimetry observations, the large collecting area of the {\em E-ELT} should be coupled with dedicated instruments. Indeed, the paucity of optical pulsation and polarisation measurements obtained so far is explained not only by the intrinsic object faintness  but also by the very limited number of such instruments offered on 8m-class telescopes, in particular for high time resolution imaging/photometry.   The most recent detection of optical pulsations from INSs have been indeed achieved through the use of guest instruments like, e.g. {\em ULTRACAM}.  While the {\em E-ELT} will certainly offer imaging and spectroscopy capabilities,  possibilities for polarimetry and high-time resolution observations are still open, being their scientific impact more and more acknowledged in the astronomical Community at large. 

The large collecting area of the {\em  E-ELT} and its large spatial resolution achievable with the AO  would be also crucial to measure proper motions and parallaxes of radio-silent INSs on a much shorter time scale through imaging astrometry.  In particular, the {\em MICADO} instrument at the {\em E-ELT} will allow to measure INS proper motions up to the LMC distance and to measure INS parallaxes for objects as distant as $\sim 1$ kpc, or more. Proper motion measurements in the optical/IR have been obtained so far for 9 INSs (half of which are radio-silent) and parallax measurements for 4 of them (mostly the radio-silent ones).  Since nearly half of the  sample of optically identified INSs consists of radio-silent objects, and many more of them are being identified through high-energy observations, optical/IR astrometry is the only way to measure their proper motions and parallaxes. Although the former can be measured in X-rays by {\em Chandra} (at least in some cases) this would require a much longer time span to cope with the still lower spatial resolution of its X-ray detector with respect to the optical ones. This clashes with the fact that {\em Chandra} has been now operational for 11 years and it will be decommissioned on the  long run. On the other hand, {\em IXO}, although more powerful in terms of collecting area than {\em Chandra}, will not provide the same spatial resolution, making X-ray astrometry no viable option to measure proper motions of INSs in the future.


\section{Discovery potentials}

Optical astronomy holds the potential of addressing, together with observations at other wavelengths, some of the fundamental issues in neutron star physics. In particular, it can allow to determine whether the neutron star luminosity evolves with the age, it can contribute to investigate the unknown properties and composition of matter at supra-nuclear densities in the neutron star interior,   the presence and composition of an atmosphere, and the properties of the neutron star  magnetic field and magnetosphere and its possible interactions with their highly-magnetised environments. Moreover, optical/IR astronomy can shed light on different INS formation scenarios, supernova explosion models, and disc formation in the post-supernova phases, which ultimately will allow to outline the long-sought "grand unification" scenario for INSs (Kaspi 2010).  Achieving these goals is possible through the optical astronomy advancement guaranteed by the {\em E-ELT}.
 
According to the magnetic dipole model, the neutron star luminosity is powered by the star rotation.  Observationally,  the multi-wavelength luminosity is found to be related to the neutron rotational energy loss rate as $L_{\lambda} \propto \dot{E}^{\alpha}$, where the index $\alpha$ depends on the wavelength (Zharikov et al. 2006; Becker 2009, Abdo et al. 2010). As a consequence, since in the magnetic dipole model $\dot{E}$ is related to the period and its period derivative, $\dot{E} \sim \dot{P}/P^{2}$, and both evolve with time, the multi-wavelength luminosity is also expected to evolve with time. The first possible measurements of luminosity evolution in neutron stars has been found through optical observations of the Crab pulsar, which showed a decrease in luminosity of a few thousandths of magnitude per year  (see Mignani 2010b and references therein). In principle, the {\em E-ELT} can carry out such measurements with a much higher statistical significance and on a broader sample of INSs, thus allowing to assess this effect more quantitatively.

Information on the neutron star interior can be obtained from the study of the cooling radiation from the surface. Indeed, several INSs, mainly the XDINSs but also the middle-aged rotation-powered pulsars, are known to emit thermal, blackbody (BB), X-ray radiation from a sizeable fraction of the neutron star surface at temperatures  $T \sim 40-100$ eV. In many of them, thermal emission, presumably from a larger and colder fraction of the neutron star surface, is also observed in the optical-UV.  The only way to accurately determine, coupled with the source distance, the neutron star surface temperature in the colder regions and, thus, to build the neutron star surface thermal map, is by characterising the optical Raileygh-Jeans  (R-J) spectrum through spectroscopy observations with the {\em E-ELT}.  Phase-resolved spectroscopy would also allow to measure the thermal spectrum evolution as a  function of the neutron star rotation phase and to better locate emission regions at different temperatures.  Building the thermal map on the neutron star surface  is the much needed step to study conductivity in the neutron star interior, hence the neutron star interior structure and composition. In particular, the bulk surface temperature for INSs older than a few Myears which are presumably too cold ($T < 20$ eV) to emit X-rays apart from hot, small, polar caps, can only be measured through {\em E-ELT} optical spectroscopy. This is of paramount importance to constrain the neutron star cooling curves which predict different decay rates for ages in excess of a few Myears and to investigate the efficiency of re-heating processes in the neutron star interior. 

The presence of an atmosphere around INSs has been for a long time speculated but it is still to be firmly demonstrated. Distortions in the BB spectrum from the neutron star surface are expected to be  induced by the presence of an atmosphere, with the effect being dependent on its degree of ionisation and its magnetic field. More noticeably,  spectral (absorption) features are expected to be produced by the interaction of the radiation field with atoms in the atmosphere. Some of these features are expected to appear also in the optical domain and, thus, can be detected through {\em E-ELT} spectroscopy of the INS optical counterparts. The detection of these features against the BB-like continuum will allow to investigate the composition of the atmosphere and its ionisation level, to measure the atmosphere density and opacity, and to measure the gravitational red-shift. Moreover, the magnetic field level in the atmosphere can be inferred through deep imaging polarimetry observations with the {\em E-ELT}  which would allow to measure the polarisation degree of the neutron star surface radiation as it passes through the atmosphere.  

The properties of the neutron star magnetic field, of its magnetosphere, and of the emission processes therein, can only be studied through coordinated multi-wavelength observations.  In particular, disentangling different spectral emission components in the optical/IR through {\em E-ELT} spectroscopy will be crucial to complete the spectrum of the INS  non-thermal emission over several decades in energy. In the optical, this is ascribed to synchrotron radiation, hence described by a power-law (PL) spectrum of varying slope, produced by relativistic particles in the neutron star magnetosphere. Locating the emission regions is possible  from the characterisation of the pulsar optical/IR light curve and from the comparison of its relative pulse  phase and width  with respect to the light curves at other energies.  A precise characterisation of the optical/IR pulsars' light curves will make it possible to search for microscopic random variations of the peak-to-peak, or peak-to-continuum, flux ratio produced by the occurrence of giant optical pulses (GOPs). These phenomena, the equivalent of the giant radio pulses (GRPs) observed in several radio pulsars, have been observed so far only for the bright ($V\sim 16.5$) Crab pulsar (Shearer et al. 2003) simultaneously in the optical and in the radio bands. This suggests that coherent and incoherent radio emission processes are linked at some level.  Thus, observations and monitoring of GOPs in other optical/radio pulsars is fundamental to verify such a link. Of course, given the small effect ($\sim 3 \%$) the search for GOPs requires, for most optical pulsars, a large photon statistics per phase bin, hence a large collecting area that only an ELT can offer. 

Spectral features of cyclotron origin can, in principle, be superimposed on the monotonic PL continuum and detected through {\em E-ELT} spectroscopy. A direct measurement of the neutron star magnetic field, usually only {\em inferred} from the spin parameters can, thus, be obtained by measuring the wavelength of these features. Moreover,  spectral breaks in the optical magnetospheric spectrum can be best studied through  {\em E-ELT} spectroscopy. In present cases where an adequate spectral coverage is available, either through multi-band photometry or spectroscopy, evidence for multiple breaks have been observed in the slope of the PL continuum from, e.g. the optical-UV to the IR and from the optical to the X-rays. Indeed, such spectral turnovers seem to be a common feature in the magnetospheric emission of rotation-powered pulsars (see, e.g. Mignani et al. 2010), which probably suggests a complex particle distribution in the neutron star magnetosphere.  This can be mapped through phase-resolved spectroscopy observations which would allow to unveil changes of the PL spectral index as a function of the spin period, thus unveiling differences in the emission process, corresponding to different projected regions of the neutron star magnetosphere close the magnetic poles.

Models of the neutron star magnetic field and magnetosphere can be tested through polarimetry measurements. While waiting for the X-ray polarimeter aboard {\em IXO}, polarimetry measurements  of INS outside the radio band are, so far, possible only in the optical/IR which is obviously also the only channel currently available for radio-silent objects. Measurements performed with the {\em E-ELT} will be important to  accurately determine, for the first time,  the level of the intrinsic polarisation of the emitted radiation, as well as to constrain the magnetic field dipole angle with respect to the neutron star spin axis (see, e.g. Mignani et al. 2007).  Particularly interesting would be the study of the INS magnetic field properties in non-stable configurations which are expected to occur in the case of the magnetars,  following the onset of a burst or a giant flare. Thanks to the {\em E-ELT}, it would be possible to measure the burst collimation  angle from the polarisation direction, measured immediately after the event, and to trace the magnetic field evolution as a function of the source state from the variation of the polarisation degree and position angle. The geometry of the neutron star magnetic field can be inferred from phase-resolved polarimetry observations.  In the case of the Crab pulsar, it is clear that both the polarisation degree and position angle are a strong function of the rotation phase,   which allows to map changes in the magnetic field properties around the magnetic poles.  Evidence of magneto-dynamic interactions between the INS and its surrounding PWN or SNR can be found from the precise determination of the polarisation direction and its alignment with the neutron star velocity vector and its spin axis (e.g. Mignani et al. 2007). Evidence of interaction, with the possible presence of jets and wisps, can be also found from the magnetic field topology in the PWNe and SNRs, which can be traced  from high spatial resolution polarisation maps.

Apart from providing an alternative way to measure INS proper motion and parallaxes, indeed {\em the 
only way} for radio-silent INSs, hence information on the INS distances and transverse velocities, optical/IR astrometry is also a very powerful tool for archaeastronomy investigations.  Although it is assumed that INSs are born in supernova explosions, it is still hotly debated whether the characteristics of the progenitor star,  the supernova dynamics, or processes occurring in the post-supernova phase, establish the ultimate fate of a massive star, driving a newborn neutron star to become a radio pulsar, a magnetar, or whatever different object it can turn into.  One of the best way to address this issue is to study stars likely to have been born from the same primordial cloud and belonging to the same generation of the INS progenitors.  Once the INS radial velocity is inferred, e.g. from 3D  modelling of an observed bow shock, the measurement of the transverse velocity makes it possible to simulate the INS orbit in the galactic potential. This allows  to localise the INS birth places by backward extrapolating its orbit for the spin-down age, and to identify the putative parental cluster (provided that the cluster orbits have been also accounted for).  The study of the  parental clusters  might then disclose the origin of the INS diversity.    In particular, it will be possible to shed light on the origin of the magnetars and of their extreme magnetic fields. A proposed scenario is that these are produced by dynamo processes in the proto-neutron star after an hyper-energetic supernova explosion of a hyper-massive ($>40 M_{\odot}$) star. {\em E-ELT} deep imaging and spectroscopy would then be able to determine the intrinsic characteristics of the cluster stars and, from them, infer those of the magnetar progenitors. Moreover, through optical/IR astrometry  with the {\em E-ELT} it will be possible to measure, on a broad sample, magnetar transverse velocities and infer the kick velocities at birth, from which one can constrain the energetic of the supernova explosion.  An alternative scenario is that the extreme magnetar magnetic fields are produced from the collapse of hyper-magnetic ($> 1000$ G)  progenitors. In this case, {\em E-ELT} spectro-polarimetry would allow to measure the magnetic fields of the parental cluster stars from the Zeeman line splitting, an effect so far measurable only for  bright ($V<14$)  and nearby stars, even with the {\em VLT}, and verify the existence of hyper-magnetic stellar populations.

The origin of the INS diversity can also reside, at least in some cases. in processes which took place right after the supernova explosion and which affected the early stages of the INS evolution like, e.g. the interaction with a  debris discs formed out of the supernova explosion.  Deep IR imaging and spectroscopy with the METIS instrument at the {\em E-ELT} will allow to detect faint debris discs around INSs, constrain their mass, size, and temperature. This is crucial to test supernova explosion models, to constrain disc/INS torque interactions, disc/INS accretion and  neutron star surface re-heating, and to investigate post-supernova planet formation around INSs. Moreover, the study of the INS optical/IR pulsations with the {\em E-ELT} would allow to isolated pulsed and DC emission components in the light curve, the latter possibly produced by a disc, as well as to infer the disc presence by searching for evidence of re-processing of  the X-ray radiation from the INS, i.e. smearing of the optical/IR light curve with respect to the X-ray one, and optical/IR-to-X-ray phase lags.


%
 
\section{Summary and conclusions}

About forty years after the identification of the first INSs, the Crab pulsar, optical astronomy of INSs is still a very active, and rapidly  evolving, field involving larger and larger parts of the Community. Optical/IR observations are important tiles to complete the multi-wavelength phenomenology picture of INSs, to address several open issues in neutron star physics, from the structure and composition of the interior, to the geometry of the magnetic field, and are crucial to understand the observed diversity among different types of INSs and, eventually, to draw a unified scheme of INS formation and evolution. While observations performed with the {\em HST} and 8m-class telescopes like the {\em VLT} have already provided important results in many respects, it is clear that more in-depth investigations are required to ensure a major step forward and to fill the gap which still exists with respect to observations at other wavelengths. This opportunity will be provided by the next generation of extremely large telescopes, like the {\em E-ELT}, but also with other similar projects, like the Thirty Meter Telescope (TMT), and the Giant Magellan Telescope (GMT), to be built in Mauna Kea (Hawaii) and in  Las Campanas (Chile), which will be operative around 2020. This will contribute to open a new era in neutron star astronomy which will count on the power of a full suite of megastructure observing facilities at all wavelengths.

\begin{acknowledgements}
The author acknowledges OPTICON for financial support for the conference participation 
\end{acknowledgements}

\end{document}